\documentclass[12pt]{article}
\usepackage{epsfig}
\begin{document}

\title{Anisotropic spectra of acoustic type turbulence}
\author{E. Kuznetsov $^{(a,b)}$  and V. Krasnoselskikh $^{(a)}$\\
\textit{\small $^{(a)}$ LPCE, 3A Avenue de la Recherche Scientifique
45071 Orleans, CEDEX 2, France} \\
\textit{\small $^{(b)}$ P.N. Lebedev Physical Institute, RAS, 53 Leninsky Ave., 119991 Moscow, Russia}
}
\date{}
\maketitle

\begin{abstract}
We consider the problem of spectra for acoustic type of turbulence generated
by shocks being randomly distributed in space. We show that for turbulence
with a weak anisotropy such spectra have the same dependence in $k$-space as
the Kadomtsev-Petviashvili (KP) spectrum: $E(k)\sim k^{-2}$. However, the
frequency spectrum has always the falling $\sim \omega^{-2}$, independently
on anisotropy. In the strong anisotropic case the energy distribution
relative to wave vectors takes anisotropic dependence forming in the large $%
k $ region the spectra of the jet type.
\end{abstract}

\medskip

PACS: 52.35.Ra, 52.35.Tc, 47.35.Rs, 47.35.Jk

\section{Introduction}

\vspace{0.5cm}

The acoustic type turbulence is ubiquitous in space and laboratory plasmas.
Typical example represents MHD turbulence (for review see \cite{SchekochihinCowley} and references therein) 
in the presence of external
magnetic field at a moderate level of $\beta $ ($\beta $ is the ratio of
kinetic plasma pressure to the magnetic pressure). When wave amplitudes are
small the turbulence can be described as an ensemble of linear
noninteracting waves with their frequencies $\omega _{k}$ and wave-vectors $%
k $. In the long-wave limit wave frequency can be considered as linear
function of a wave number: $\omega _{k}\sim k$. In the thermodynamic
equilibrium limit these waves are distributed according to the Jeans law so
that the energy of each wave $\varepsilon _{k}=T$ where $T$ stands for the
"wave's" temperature. When the wave amplitudes increase (with the growth of
the energy source) we first come to the regime of weak turbulence. The
nonlinear effects are weak in comparison with linear wave dispersion but
statistical characteristics of the system change significantly. In the weak
turbulence regime, in addition to the thermodynamic distribution solution,
it emerges the additional solution having the spectrum of the Kolmogorov
type: 
\[
\varepsilon (k)=C\left( \rho Pc_{s}\right) ^{1/2}k^{-7/2} 
\]%
where $P$ is the energy flux toward small scales, $\rho $ the density, $%
c_{s} $ the sound velocity and $C$ is the dimensionless constant of the
order of one. In the spherical normalization this energy distribution reads
as follows 
\begin{equation}
E(k)=4\pi k^{2}\varepsilon _{k}=4\pi C\left( \rho Pc_{s}\right)
^{1/2}k^{-3/2}.  \label{ZS}
\end{equation}
This spectrum was found in 1971 by Zakharov and Sagdeev \cite%
{ZakharovSagdeev} as the exact solution of the wave kinetic equation for
acoustic waves in isotropic medium, when in the long-wave region the
dispersion relation is linear $\omega _{k}\approx c_{s}k$ (for details, see \cite{ZLF}). In this case the
criterion of weak turbulence that is determined as the weakness of the
nonlinearity with respect to the wave dispersion, is written as 
\begin{equation}
\frac{\Delta \rho }{\rho }\ll k^{2}\Lambda ^{2}  \label{weakcrit}
\end{equation}%
where $\Delta \rho $ is the fluctuating part of the density $\rho ,$ and $%
\Lambda $ is the dispersion length \footnote{%
In the case of three-wave interacting waves of different branches the
criterion of weak turbulence consists in smallness of the inverse nonlinear
time defining by the kinetic equations in comparison with the maximal growth
rate of the corresponding decay instability for the monochromatic wave. The
latter in such system represents the inverse time of randomization (see, e.g. \cite{kuzn2001})}. For
larger amplitudes when the nonlinear effects become comparable or larger the
wave dispersion this criterion (\ref{weakcrit}) breaks down. As it is well
known in gas dynamics, in this case the main nonlinear effect for acoustic
waves is nothing else but the wave breaking that results in the formation of
shocks (or discontinuities). It is well known also that this process in
compressible flows can be treated in terms of the formation of folds in the
classical catastrophe theory \cite{arnold}. In the gas-dynamic case,
breaking areas can be completely characterized using the mapping defined by
the transition from the usual Eulerian to the Lagrangian description. A
vanishing of the Jacobian of the mapping corresponds to the emergence of a
singularity for the spatial derivatives of the velocity and density of the
fluid. Physical meaning of this effect corresponds to the intersection of
particle trajectories. In the general situation first time the Jacobian
vanishing happens in one isolated point. In collisionless plasma this
process can continue and leads further to form the multi-flow region
expanding in the transverse (relative to the main flow) direction according
to the following scaling law $R_{\perp }$ $\sim \sqrt{t_{0}-t}$, where $%
t_{0} $ stands for the first moment when Jacobian $J$ turns to zero, while
the region width (in the longitudinal direction, along the main flow)
increases more slowly, $R_{\parallel }\sim (t_{0}-t)^{3/2}$ (see, e.g. \cite%
{krasn, ShandarinZeldovich}). Thus, the result of the breaking consists in the formation of
structures in the form of pancakes with very different characteristic
spatial scales along the flow and in transverse directions. In optics quite
similar structures are called caustics \cite{LandauLifshitz}. The simplest way to
represent such structures is to consider them as disks (singular manifolds)
on which the density undergoes jumps vanishing at the disk boundary. It is
worth noting that in classical hydrodynamics these regions are considered as
"forbidden" where the solution can not be constructed anymore in a well
defined way, that means as a single value solution. There are special
procedures of construction of so called "shock type" solution, see the book
of Whitham \cite{whitham} for more detail. However, in collisionless plasma physics the
detailed description is written in terms of velocity distribution function
that satisfies to Vlasov equation in the phase space $({\bf p_{i}},%
{\bf r_{i}})$. For this function the regions where the
hydrodynamic solution becomes multi-valued and poorly determined are the
areas where the distribution function undergoes the transition from the one
single peak distribution to the one having three peaks, however there is no
any crucial break in the phase space. From the other hand the spatial
derivatives of some characteristics have infinite gradients also, but this
effect can be explicitly described in the frame of the same Vlasov equation
for collisionless plasmas. Formation of simple waves making use of such a
description was considered by Gurevich and Pitaevskii \cite{GurevichPitaevskii}.

In weak turbulence waves are assumed to be randomly distributed with a weak
correlation between waves because of weak nonlinear interaction. The process
of breaking is purely coherent and, respectively, the expanding caustics
should be treated as coherent elementary entities of wave turbulence. We
will call the turbulence where this effect becomes important a moderately
strong turbulence if the density of caustics is still small enough to
neglect their intersections that are supposed to be rare. The caustics can
be described if their centers and orientations are determined. We will
assume their distribution in space to be random. From the other hand, the
jumps as the density singularities are known to result in a power-law tails
in the short-wavelength part of the turbulence spectrum. This idea was first
proposed by Phillips \cite{phillips} and allowed him to determine the
water-wave turbulence spectrum in the presence of whitecaps, i.e., of the
singularities on the fluid surface. Later the very same idea was developed
by Saffman \cite{saffman} to determine the isotropic spectrum of two-dimensional
hydrodynamic turbulence at high Reynolds numbers (for  details see \cite{kuzn,KHNR07}). Kadomtsev and Petviashvili
(KP) \cite{KP} suggested that the acoustic turbulence  can be
considered as a randomly distributed set of shocks. For the isotropic case
they found the spectrum of the energy distribution, now known as the KP
spectrum, 
\begin{equation}
E(k)~\sim ~k^{-2}.  \label{kp}
\end{equation}%
The goal of our paper is to determine how the KP spectrum is modified in the
presence of strong enough anisotropy in plasma. Such an anisotropy can be
entirely inherent, either due to the mean magnetic field or because of the
anisotropy of turbulence source/pumping. It is worth noting that
in the weak-turbulence regime, when the wave dispersion of acoustic waves  
can be neglected, the angular distribution of the spectrum repeats 
the   anisotropy of the pumping because of three-wave resonant conditions, $\omega_k=\omega_{k_1}+\omega_{k_2}$ and
${\bf k}={\bf k_1}+{\bf k_2}$. In this case the additional anisotropy of the spectra can 
appear either due to weak dispersion \cite{LvovFalkovich, ZLF} 
or because of nonlinear renormalizations to the wave kinetic equation \cite{LLNZ}.
For weak MHD turbulence, in both low and high beta plasmas, the situation is familiar to
the weak acoustic turbulence in isotropic media for the nonlinear interacting fast magnetoacoustic waves which  dispersion is mainly defined by modulus of the wave vector. However, at $\beta\ll 1$  slow magnetoacoustic waves undergo strong anisotropy in both dispersion and nonlinear interaction due to the external magnetic field. As it was shown in the paper \cite{kuzn72} by one of the authors, it results in appearance of anisotropic Kolmogorov-type spectra with power dependences relative to both longitudinal and transverse projections of the wave vector. Such character of the spectra can be established also for the interaction of Alfvenic and 
slow  magnetoacoustic waves at low beta plasma \cite{kuzn2001}.   
For $\beta\gg 1$ slow magnetoacoustic waves have the same dispersion as the Alfvenic waves. This degeneracy also changes  significantly the nonlinear interaction  between waves and leads in the weak-turbulence regime  to the spectrum of the Kraichnan-Iroshnikov type \cite{IroshnikovKraichnan} (for details see \cite{GNNP,SchekochihinCowley}).

In the present paper, for moderately strong acoustic turbulence, we study  how the angular ordering of 
shocks can change the angular structure of the spectrum. We show that
for the strong anisotropy the spectrum  has the jet-type behavior with
power increasing  along the jet with the same exponent as for the isotropic KP spectrum 
and the falling-off dependence in transverse direction   $\sim k_{\perp}^{-5}$. The latter  
originates from the contribution from boundaries of caustics (disks). It is necessary to mention
that for two-dimensional acoustic turbulence the phenomenon of the jet-type spectra generated by shocks was observed first time in the numerical experiments
\cite{FalkovichMeyer}. However, comparably small spatial resolution could not allow the authors \cite{FalkovichMeyer}
to treat the fine structure of jets. 

The second objective of our work is
to show how the presence of the KP spectrum and its relative role can be
evaluated making use of real experimental data ( for instance, from
spacecraft data for solar wind turbulence) in the presence of weak
turbulence tails falling off more slowly than the KP spectrum. This
question is very important because in the case when weak and strong
turbulent components co-exist, the KP component of the spectrum can provide
a quantitative measure of the relative role of coherent structures for
moderate acoustic type turbulence. Hereafter we shall also show that the
spectral index of the isotropic KP spectrum (\ref{kp}) as a function of the
wave-vector does not depend on space dimension. It is worth noting that the
Fourier transform of the density single point auto-correlation function
dependence upon time shows the same power-law spectral index in time domain
(in frequency) as in the spatial domain (spectrum in $k$) \ref{kp}.

\section{Isotropic spectra}

Let us consider the contribution of shocks to the frequency spectrum, i.e.,
the energy distribution dependence on frequency. To find the spectrum one
should calculate the auto-correlation function for the  density $\rho (t)$
measured at some point $\mathbf{r}_{0}$ as a function of time: $K(\tau
)=\left\langle \rho (t+\tau )\rho (t)\right\rangle $ where the angular
brackets stand for time averaging of the (mass) density $\rho (t)$ and then
carry out the Fourier transform $K(\tau),$%
\[
K_{\omega }=~\int_{-\infty }^{\infty }K(\tau )e^{i\omega \tau }d\tau. 
\]%
Here  the density distribution is assumed to be  homogeneous so that $K_{\omega }$ does
not depend on $\mathbf{r}_{0}$. It is worth noting that defined in such a
way, $K_{\omega }$ coincides, up to constant factor, with the energy
spectrum in the frequency domain: 
$$
E_{\omega }=\frac{c_{s}^{2}}{\rho _{0}}K_{\omega }. 
$$
In the weak-turbulence approximation, in an isotropic case $E_{\omega }$ can
be expressed in terms of the spectrum $\varepsilon (k)$ by means of the
relation 
\begin{equation}
E_{\omega }=~\frac{4\pi \omega ^{2}}{c_{s}^{3}}~\varepsilon (k)
\label{weakomega}
\end{equation}%
where $k=\omega /c_{s}$. This formula is the result of integration of the
energy spectral density in the $k-\omega $ representation 
$E_{k\omega }=\varepsilon (k)~\delta (\omega -\omega _{k})$
with respect to $\mathbf{k}.$ In the weak turbulence regime the spectral density $%
E_{k\omega }$ has the $\delta $-function dependence upon frequency
indicating that the wave ensemble is weakly nonlinear. In the presence of
shocks, i.e. for the strong turbulence regime, such relations are not valid
any more.

Our aim now is to determine the contribution of shocks associated with the density
jumps to the $E_{\omega }$ spectrum. To achieve this one should take into
account that at the instant $t_{i}$ of jump passage through the measurement
point $\mathbf{r}_{0}$ the first derivative ${\partial \rho }/{\partial t}$
is proportional to $\delta (t-t_{i})$, i.e., 
\begin{equation}
\frac{\partial \rho }{\partial t}=\sum\limits_{i}\Delta \rho _{i}\delta
(t-t_{i})+\mbox{regular terms}.  \label{jump-t}
\end{equation}%
Assuming that density jumps $\Delta \rho _{i}$ and crossing times $t_{i}$
are random quantities one can calculate the contribution of these
singularities in Eq. (\ref{jump-t}) to the spectrum. The Fourier transform
of the contribution associated with these terms can be written as follows: 
\begin{equation}
\rho _{\omega }=\frac{i}{2\pi \omega }\sum\limits_{i}~\Delta \rho
_{i}e^{-i\omega t_{i}}.  \label{7}
\end{equation}%
Here, 
\[
\rho _{\omega }=\int_{-\infty }^{\infty }\rho (t)e^{i\omega
t}dt,\,\,\,\,\rho (t)=\frac{1}{2\pi }\int_{-\infty }^{\infty }e^{-i\omega
t}\rho _{\omega }d\omega . 
\]%
To find the $E_{\omega }$ spectrum, one should square the absolute value of (%
\ref{7}) and average the resulting expression. The averaging over crossing
times $t_{i}$ yields 
\begin{equation}
E_{\omega }=\frac{c_{s}^{2}}{2\pi \rho _{0}\tau }\langle |\rho _{\omega
}|^{2}\rangle =\frac{c_{s}^{2}\nu }{2\pi \rho _{0}\omega ^{2}}\overline{%
(\Delta \rho )^{2}},  \label{omega}
\end{equation}%
where $\nu =N/\tau $ is the jump crossings frequency, here $N$ is the number
of discontinuities met during the averaging time $\tau $, and $\overline{%
(\Delta \rho )^{2}}$ is the average value of $(\Delta \rho )^{2}$.

The very same approach can be applied for finding the spatial spectrum of
acoustic turbulence for the one dimensional case ($D=1)$ when instead of (%
\ref{jump-t}) we have 
\begin{equation}
\frac{\partial \rho }{\partial x}=\sum\limits_{i}\Delta \rho _{i}\delta
(x-x_{i})+\mbox{regular terms}  \label{jump1D}
\end{equation}%
where $x_{i}$ are the positions of jumps along $x$-axis. Hence we get the
following $1D$ spectrum $E(k)$: 
\begin{equation}
E_{1}(k)=\frac{n_{1}c_{s}^{2}}{2\pi \rho _{0}k^{2}}~\overline{(\Delta \rho
)^{2}}.  \label{oneD}
\end{equation}%
Here, $n_{1}$ is the number density of shocks per unit length, $\rho _{0}$
is the mean density of medium (per unit length), and $\overline{(\Delta \rho
)^{2}}$ is the mean-square of the density jump at the discontinuity.

This calculation can be considered in isotropic three dimensional case as
the estimate of the correlation function along any chosen straight line. Now
we shall make use of it to obtain the three-dimensional (3D) isotropic
spectrum (\ref{oneD}). To this end one should notice that the density
correlation function for isotropic turbulence, 
$\phi (y_{1})=\langle \rho (x_{1}+y_{1},x_{2},x_{3})\rho
(x_{1},x_{2},x_{3})\rangle$, 
has the Fourier spectrum (with respect to only one single variable $y_{1}$!)
that coincides, to within a factor, with Eq. (\ref{oneD}): 
\begin{equation}
\phi _{k}=\frac{N_{1}}{2\pi k^{2}}~\overline{(\Delta \rho )^{2}},
\label{phi}
\end{equation}%
where $N_{1}$ in this case should be considered as the mean linear density
of discontinuities. This correlation function $\phi _{k}$ is related to the
three-dimensional Fourier spectrum 
\[
\Phi (|\mathbf{k}|)=\int \phi (\mathbf{r})e^{-i\mathbf{k\cdot r}}d\mathbf{r} 
\]%
by the following formula 
\[
\phi _{k_{1}}=\int \Phi (|\mathbf{k}|)d\mathbf{k_{\perp }}=\pi
\int_{k_{1}^{2}}^{\infty }\Phi (s)ds^{2} 
\]%
where $\mathbf{k}_{\perp }$ is the component of the wave vector $\mathbf{k}$
perpendicular to the $x$-axis. This allows one to obtain
differentiating this equality with respect to $k_{1}$ the following
relation: 
\[
\Phi (k)=-\frac{1}{2\pi k}\frac{d\phi _{k}}{dk}. 
\]%
Then, substituting Eq. (\ref{phi}), one can find the following expression
for the spectrum $E_{3}(k)$: 
\begin{equation}
E_{3}(k)=\frac{2N_{1}c_{s}^{2}}{\pi \rho _{0}k^{2}}~\overline{\Delta \rho
^{2}}.  \label{threeD}
\end{equation}%
The same approach is applicable in two-dimensional case. It is a little more
difficult technically because it requires to solve the integral Abel
equation (e.g. compare with \cite{saffman, kuzn}). However, making calculations one can find that the spectrum
dependence upon $k$ has the same spectral index as in one- and three-
dimensional cases: 
\[
E_{2}(k)\propto k^{-2}. 
\]

According to our knowledge spectrum (\ref{oneD}) was first obtained by
Burgers in \cite{burgers} and, its generalization for multi-dimensional
situation was found (\ref{threeD}) by Kadomtsev and Petviashvili \cite{KP}.
In the next section we show how to generalize it taking into account the
effect of anisotropy.

\section{Anisotropic KP spectra}

The analysis of anisotropic situation consists in taking account of two
geometric factors. One is connected with the anisotropy of the caustics
orientations, another with the anisotropy due to the "emission" diagram of
each caustic that can be considered as the source of the wave spectrum. For
the sake of simplicity and without loss of generality we can consider that
each single caustic has the form of a disk. This simplification is based on
important property of these objects, namely, during their evolution most of
time their relative size in the perpendicular direction is much larger than
along the parallel axis.

Let us consider a single caustic with radius $R_{i}$ perpendicular to the $x$
-axis and centered at the point $\mathbf{r}_{0}=(x_{0},\mathbf{r}_{\perp 0})$%
. To consider this effect equation (\ref{jump1D}) should be replaced by the
following one: 
\begin{equation}
\frac{\partial \rho }{\partial x}=\Delta \rho (|\mathbf{r}_{\perp }-\mathbf{r%
}_{\perp 0}|)\delta (x-x_{0})+\mbox{regular
~ terms}.  \label{jump3D}
\end{equation}%
Here, $\Delta \rho (r_{\perp })$ is considered as a continuous
cylindrically-symmetric function of $r_{\perp }$ that vanishes at the disk
boundary $r_{\perp }=R$ $(\Delta \rho (R)=0)$ and remains zero outside the
disk.

Then, the Fourier transform of the terms corresponding to singular part of
Eq. (\ref{jump3D}) is given by the integral 
\[
\rho _{\mathbf{k}}=-\frac{i}{k_{x}^{{}}}e^{-i\mathbf{kr}_{0}}\int_{r_{\perp
}\leq R}~\Delta \rho (r_{\perp })e^{-i\mathbf{k}_{\perp }\mathbf{r}_{\perp
}}d\mathbf{r}_{\perp }=-\frac{2\pi i}{k_{x}^{{}}}e^{-i\mathbf{kr}%
_{0}}\int_{0}^{R}~r_{\perp }\Delta \rho (r_{\perp })J_{0}(k_{\perp }r_{\perp
})dr_{\perp }, 
\]%
where $\mathbf{k}=(k_{x},\mathbf{k}_{\perp })$ and $J_{0}(k_{\perp }r_{\perp
})$ is the Bessel function. This is the contribution from one single
singularity. The total contribution from all discontinuities can be found as
the sum: 
\[
\rho _{\mathbf{k}}=-2\pi i\sum\limits_{\alpha }\frac{e^{-i\mathbf{kr}%
_{\alpha }}}{\mathbf{kn}_{\alpha }}\int_{0}^{R_{\alpha }}~\Delta \rho
(r_{\perp })~r_{\perp }J_{0}(k_{\perp \alpha }r_{\perp })dr_{\perp }. 
\]%
Here, $\mathbf{n}_{\alpha }$ is the normal unit vector to the discontinuity $%
\alpha $, $\mathbf{r}_{\alpha }$ are the disk center coordinates and $%
k_{\perp \alpha }$ is the transverse projection of the wave vector $\mathbf{k%
}$ to the disc plane ($k_{\perp \alpha }^{2}=k^{2}-(\mathbf{kn}_{\alpha
})^{2}$). It is worth noting here that  the anisotropic characteristics of the spectrum is
related to the anisotropy of the distribution of unit vectors $\mathbf{n}%
_{\alpha }$.

To find the spectrum of turbulence, one should average $|\rho _{k}|^{2}$
over all random variables. It is natural to assume that the coordinates $%
\mathbf{r}_{\alpha }$ of centers of caustics are distributed uniformly, the
averaging over these variables results in 
\begin{equation}
\overline{|\rho _{k}|^{2}}=4\pi ^{2}N\left\langle \left\vert \frac{1}{%
k_{\Vert }^{{}}}\int_{0}^{R}~\Delta \rho (r_{\perp })r_{\perp
}J_{0}(k_{\perp }r_{\perp })dr_{\perp }\right\vert ^{2}\right\rangle .
\label{aver1}
\end{equation}%
Here, $k_{\Vert }^{{}}\equiv \mathbf{kn}$, $N$ is the density of
discontinuities and the angular brackets stand for the averaging over
characteristic sizes $R$ and angles.

Our major interest here is in evaluation of the short-wavelength asymptotic
behavior of the spectrum found (\ref{aver1}), i.e., $k\overline{R}\gg 1$,
where $\overline{R}$ is the characteristic disc radius. Thus, if $k_{\perp }%
\overline{R}\gg 1$, then the expression inside the integral in Eq. (\ref%
{aver1}) represents rapidly oscillating function (due to the Bessel function 
$J_{0}(k_{\perp }r_{\perp })$) and therefore the integral can be evaluated
by means of the stationary phase method. Using the relation
\[
xJ_{0}(\alpha x)=\frac{1}{\alpha }\frac{d}{dx}\left( xJ_{1}(\alpha x)\right)
, 
\]%
and integrating by parts we have 
\[
\int_{0}^{R}~\Delta \rho (r_{\perp })r_{\perp }J_{0}(k_{\perp }r_{\perp
})dr_{\perp }=-\frac{R^{{}}}{k_{\perp }}\int_{0}^{1}J_{1}(k_{\perp }Rx)~x%
\frac{d}{dx}\Delta \rho (Rx)dx 
\]%
where the property $\Delta \rho (R)=0$ is used. Having in mind that the
major input to the integral comes from the vicinity of the boundary at $r_{\perp
}=R$ we can use the asymptotic expression for the Bessel function at large $%
k_{\perp }R$ 
\[
J_{1}(z)\simeq \sqrt{\frac{2}{\pi z}}\cos \left( z-\frac{3\pi }{4}\right) , 
\]%
and obtain the following asymptotic estimate of the integral: 
\[
-\sqrt{\frac{2R}{\pi }}\frac{^{1}}{k_{\perp }^{3/2}}\int_{0}^{1}\cos \left(
k_{\perp }Rx\right) ~x^{1/2}\frac{d}{dx}\Delta \rho (Rx)dx\sim -\sqrt{\frac{%
2R}{\pi }}\frac{\sin \left( k_{\perp }R\right) }{k_{\perp }^{5/2}}%
^{{}}\left. \frac{d}{dr_{\perp }}\Delta \rho (r_{\perp })\right\vert
_{r_{\perp }=R}. 
\]%
where the main contribution to the integral comes explicitly from the
boundary $r_{\perp }=R$.

Hence the resulting spectrum $\tilde{\epsilon}(\mathbf{k})$ (still prior to
angular averaging!) can be written as follows 
\begin{equation}
\tilde{\epsilon}_{1}(\mathbf{k})=\frac{4\pi nc_{s}^{2}}{\rho _{0}}\frac{%
\left\langle \Psi \right\rangle }{k_{\perp }^{5}k_{\parallel }^{2}},\,\,\,\Psi =R\left( \left. \frac{d}{dr_{\perp }}\Delta \rho (r_{\perp
})\right\vert _{r_{\perp }=R}\right) ^{2}. 
\label{epsilon1}
\end{equation}%
It is worth noting that the stationary phase method used for the evaluation
of the integral in Eq. (\ref{aver1}) is applicable at almost all angles $%
\theta $ ($\theta $ is the angle between the vectors $\mathbf{k}$ and $%
\mathbf{n}$) but not in two narrow cones $\theta \leq \vartheta _{0}=\left( k%
\overline{R}\right) ^{-1}$and $\pi -\theta \leq \vartheta _{0}$. Inside
these cones one can find that the integral can be considered to be
independent on $k$ (suggesting $\cos \left( k_{\perp }Rx\right) \approx 1$). The
spectrum $\tilde{\epsilon}(\mathbf{k})$ inside these cones is then given by 
\begin{equation}
\tilde{\epsilon}_{2}(\mathbf{k})\approx \frac{4\pi nc_{s}^{2}}{\rho _{0}}%
\frac{\left\langle \Gamma ^{2}\right\rangle }{k_{\Vert }^{2}},\,
\label{epsilon2}
\end{equation}%
where 
\[
\Gamma =\int_{0}^{R}~\Delta \rho (r_{\perp })r_{\perp }dr_{\perp }. 
\]%
From Eq. (\ref{epsilon1}) one can see that the spectrum $\tilde{\epsilon}%
_{1}(\mathbf{k})$ contains three singularities, namely, at angles $\theta $
close to $0,\pi $ and $\pi /2$. For angles close to the cone $\theta \approx
\left( k\overline{R}\right) ^{-1}$ and $\pi -\theta \approx \left( k%
\overline{R}\right) ^{-1}$expression (\ref{epsilon1}) matches Eq.(\ref%
{epsilon2}). For angles close to $\pi /2$, in Eq. (\ref{epsilon1}) an
additional effect should be taken into account, namely, the bending of the
caustics. If $a$ is the characteristic value of bending, Eq. (\ref{epsilon1}%
) is valid in the region $|\theta -{\pi }/{2}|>(ka)^{-1}$.

Distributions (\ref{epsilon1}) and (\ref{epsilon2}) allow one to carry out
the averaging over angular distribution of caustics normals, if their
distribution is anisotropic, i.e., calculate the angular dependence of the
energy spectrum $E(\mathbf{k})=k^{2}\overline{\tilde{\epsilon}(\mathbf{k})}$
in short enough scales $\left( k\overline{R}\right) \gg 1$.

First, let us check that in the isotropic case the KP spectrum (\ref{threeD}%
) follows from the above calculations. In this case averaging over the
angles corresponds to the integration of Eqs. (\ref{epsilon1}) and (\ref%
{epsilon2}) with respect to $\theta $. The integration of expression (\ref%
{epsilon2}) near poles $\theta =0,\pi $ gives 
\begin{equation}
E_{2}(k)=4\pi k^{2}\int_{0}^{\theta _{0}}\tilde{\epsilon}_{2}(k)\theta
d\theta =\frac{16\pi ^{2}nc_{s}^{2}}{\rho _{0}k^{2}}\frac{\left\langle
\Gamma ^{2}\right\rangle }{\overline{R}^{2}}.  \label{small}
\end{equation}%
In the integration of Eq. (\ref{epsilon1}) over angles, the main
contribution to the spectrum comes from the angles close to $0,\pi $ and $%
\pi /2$, where the spectrum (\ref{epsilon1}) has singularities. For $\theta
\rightarrow 0$ (for $\theta \rightarrow \pi $), the integration is cut off
at angles $\theta _{k}=\vartheta _{0}$ ( at $\theta _{k}=\pi -\vartheta _{0}$%
), and for $\theta \rightarrow {\pi }/{2}$ it is cut off at angles $|{\pi }/{%
2}\pm \theta _{k}|\approx ({ka)}^{-1}$. As the result of averaging of the
expression (\ref{epsilon1}) we get that the main contributions coming from
the angles near the cone: 
\begin{equation}
E_{1}(k)\approx \frac{16\pi ^{2}nc_{s}^{2}}{\rho _{0}k^{2}}\frac{\overline{R}%
^{3}\left\langle \Psi \right\rangle }{3}.  \label{large}
\end{equation}%
The spectrum $E(k)$ is given by the sum of (\ref{small}) and (\ref{large}), 
\begin{equation}
E(k)\approx \frac{16\pi ^{2}nc_{s}^{2}}{\rho _{0}k^{2}}\left( \frac{%
\left\langle \Gamma ^{2}\right\rangle }{\overline{R}^{2}}+\frac{\overline{R}%
^{3}\left\langle \Psi \right\rangle }{3}\right) ,  \label{kp1}
\end{equation}%
that has the same dependence on $k$ as the KP spectrum (\ref{threeD})
obtained in a little different way from conventional consideration.

Our above performed analysis of spectral angular dependence upon the angle
that is applicable first of all to one single caustic leads to the
conclusion that the "emission diagram" of it represents narrow cones around $%
\theta =0,\frac{\pi }{2},\pi $. The angular width of these cones is
proportional to $\left( k\overline{R}\right) ^{-1}$ around $\theta =0,\pi $
and to $({ka)}^{-1}$ around $\theta =\pi /2$. In the case of anisotropic
distribution of caustics normals the angular dependence of the spectrum will
simply reproduce this angular dependence if the width $\Delta \theta $ of
this angular distribution is sufficiently larger than the width of these
cones. This can be resumed as follows, if the angular distribution of
caustics normals is wide enough so that $\Delta \theta >\vartheta _{0}$ then
after averaging of $\tilde{\epsilon}(\mathbf{k})$ one should get the same
dependence of the spectrum upon $k$ at large enough $k$ as for the isotropic
KP\ spectrum, i.e. $\sim k^{-2}$ with the degree of the anisotropy exactly
the same as the distribution of caustics normals. From the other hand if the
angular distribution of caustics normals is sufficiently narrow , e.g., if
all falling shock fronts are oriented almost unidirectionally (this can be
caused, e.g., by pumping or initial/boundary conditions), then the spectrum
will have a sharp peak in this direction. If the width $\Delta \theta $ of
angular distribution is narrower than $\vartheta _{0}$, i.e., if $\Delta
\theta <\vartheta _{0}$, then the spectrum $E(k,\theta )$ up to the
multiplier $k^{2}$ will repeat the distribution given by Eqs. (\ref{epsilon1}%
) and (\ref{epsilon2}), namely, the distribution in $k$-space will have the
form of the \textit{jet} . In the cone $\theta <\vartheta _{0}$ the spectrum
has a maximum with fall off at large $k$ in accordance with (\ref{epsilon1}%
), i.e $\sim k_{\Vert }^{-2}$ like the KP spectrum (\ref{threeD}). At larger
angles, $\theta >\vartheta _{0}$ the spectrum $E(k,\theta )$ will rapidly
decrease in the transverse direction proportionally to $k_{\perp }^{-5}$.

Note, however, that this asymptotic behavior is intermediate, because $%
\vartheta _0=(k\overline{R})^{-1}$ decreases with increasing $k$. For this
reason, when averaging over angles, singularities in (\ref{epsilon1}) become
essential for $\theta \to 0$, and, starting from certain $k=k^{*}$, the
spectrum will decrease as $k^{-2}$ with increasing $k$. The angular width of
the spectrum will be of order $\Delta \theta$ .

\section{Acknowledgments}

The work of E.K. was partially supported by the Russian Foundation for Basic
Research (grant No. 06-01-00665), by the Council for the State Support of
the Leading Scientific Schools of Russia (grant No. NSH-4887.2008.2) and by Poste Rouge fellowship of
French National Centre of the Scientific Research.


\begin{thebibliography}{9}

\bibitem{SchekochihinCowley} A. A. Schekochihin and S. C. Cowley,
{\it Turbulence and magnetic fields in astrophysical plasmas}, 
in: Magnetohydrodynamics: Historical Evolution and Trends, S. Molokov, R. Moreau, and H. K. Moffatt, Eds.
(Berlin: Springer, 2007), 85.

\bibitem{ZakharovSagdeev} V.E. Zakharov and R.Z. Sagdeev, 
Sov. Phys. Dokl. \textbf{15}, 439 (1971).

\bibitem{ZLF}V. Zakharov, V. L'vov, and G. Falkovich, {\it Kolmogorov Spectra of Turbulence} (Springer-Verlag, Heidelberg, 1992).  

\bibitem{kuzn2001}E.A. Kuznetsov,  Zh. Eksp. Teor. Fiz.
{\bf 120}, 1213 (2001) [JETP, {\bf 93}, 1052-1064 (2001)].

\bibitem{arnold} V.I. Arnold, {\it Catastrophe Theory} (Znanie, Moscow, 1981;
Springer, Berlin, 1986); {\it Mathematical Methods of Classical Mechanics}, 3rd
ed. (Nauka, Moscow, 1984; Springer, New York, 1989).

\bibitem{krasn} V.V. Krasnoselskikh, Sov. Phys. JETP {\bf 62},
282 (1985).

\bibitem{ShandarinZeldovich} S.F. Shandarin and Ya.B. Zeldovich, Rev. Mod. Phys. 
{\bf 61}, 185 (1989).

\bibitem{LandauLifshitz}L.D. Landau and E.M. Lifshitz, {\it The Classical Theory of Fields},
London/Reading MA: Pergamon/Addison-Wesley, 1959. 

\bibitem{whitham}G.B. Witham, {\it Linear and Nonlinear Waves}, Wiley Interscience, New York, 1974.

\bibitem{GurevichPitaevskii}A.V. Gurevich and L.P. Pitaevskii, "Nonlinear Dynamics of Rarefied Plasmas and Ionospheric Aerodynamics", in: {\it Reviews of Plasma Physics}, Volume 10, p.1, 
Ed. M. A. Leontovich,  Consultants Bureau, New York, 1986. 

\bibitem{phillips} O.M. Phillips, J. Fluid Mech. \textbf{4}, 426 (1958).

\bibitem{saffman} P.G. Saffman, Stud. Appl. Maths, \textbf{50}, 49 (1971).

\bibitem{kuzn}E.A. Kuznetsov, Pis'ma Zh. Eksp. Teor. Fiz. {\bf 80}, 92 (2004) [JETP Letters {\bf 80}, 83 (2004)].

\bibitem{KHNR07}E.A. Kuznetsov, V. Naulin, A.H. Nielsen, and J.J. Rasmussen, Phys Fluids {\bf 19}, 105110 (2007).

\bibitem{KP} B.B. Kadomtsev and V. I. Petviashvili, Dokl. Akad. Nauk SSSR 
\textbf{208}, 794 (1973) [Sov. Phys. Dokl. \textbf{18}, 115 (1973)].

\bibitem{LvovFalkovich}V.S. L'vov, and G.E. Falkovich, Sov. Phys. JETP {\bf 53}, 299 (1981). 

\bibitem{LLNZ}V.S. L'vov, Yu. L'vov, A.C. Newell and V. Zakharov, Phys. Rev. E {\bf 56}, 390 (1997). 

\bibitem{kuzn72}E.A. Kuznetsov,  Zh. Eksp. Teor. Fiz., {\bf 62}, 584 (1972) [Sov. Phys. JETP {\bf 62}, {\bf 35}, 310 (1972)].

\bibitem{IroshnikovKraichnan} R.S. Iroshnikov, Astron Zh. {\bf 40}, 742 (1963) [Sov. Astron. {\bf 7}, 566 (1964)];
R.H.  Kraichnan, Phys. Fluids {\bf 11}, 945 (1965).
\bibitem{GNNP} S. Galtier, S.V. Nazarenko, A.C. Newell and  A. Pouquet, J. Plasma Phys., {\bf 63}, 447 (2000);
S.V. Nazarenko, A.C. Newell, and S. Galtier,  
Physica D, {\bf 152-153}, 646 (2001).

\bibitem{FalkovichMeyer}G. Falkovich and M. Meyer, Phys. Rev. E {\bf 54}, 4431 (1996).


\bibitem{burgers} J.M. Burgers, Lecture in Cal. Inst. Tech. (1951)
(unpublished).


\end{thebibliography}
\end{document}